\documentclass[showpacs,superscriptaddress,twocolumn,amsmath,aps]{revtex4}
\usepackage{graphicx,color}
\usepackage{bm}
\usepackage[hypertex]{hyperref}

\newcommand{\be}{\begin{equation}}
\newcommand{\ee}{\end{equation}}
\newcommand{\bea}{\begin{eqnarray}}
\newcommand{\eea}{\end{eqnarray}}
\newcommand{\bsube}{\begin{subequations}}
\newcommand{\esube}{\end{subequations}}

\newcommand{\Eq}[1]{Eq.\,(\ref{#1})}

\usepackage{amsfonts}
\usepackage{amsmath}
\usepackage{amssymb}
\usepackage{graphicx}
\usepackage{type1cm}        
\usepackage{times}          
\usepackage{bm}
\usepackage{CJK}            
\usepackage{color}                      

\newcommand{\alp}{\alpha}

\newcommand{\gam}{\gamma}

\newcommand{\eps}{\epsilon}
\newcommand{\vep}{\varepsilon}

\newcommand{\lam}{\lambda}

\newcommand{\ome}{\omega}
\newcommand{\Gam}{\Gamma}
\newcommand{\Del}{\Delta}

\newcommand{\beq}{\begin{equation}}
\newcommand{\eeq}{\end{equation}}
\newcommand{\beqn}{\begin{eqnarray}}
\newcommand{\eeqn}{\end{eqnarray}}
\newcommand{\bsub}{\begin{subequations}}
\newcommand{\esub}{\end{subequations}}
\newcommand{\iny}{{\infty}}
\newcommand{\re}{\nonumber\\}
\newcommand{\ket}[1]{{\left| #1 \right\rangle }}

\newcommand{\LL}{\mathcal {L}}

\newcommand{\cdg}{c^\dagger}
\newcommand{\ddg}{d^\dagger}
\newcommand{\fdg}{f^\dagger}

\newcommand{\tr}{{\rm Tr}}

\begin{document}
\begin{CJK*}{GBK}{song}

\title{ Probing the existence and dynamics of Majorana fermion
          via transport through a quantum dot  }

\author{Yunshan Cao}
\affiliation{School of Physics, Peking University, Beijing 100871,China}
\author{Peiyue Wang}
\affiliation{Department of Physics, Beijing Normal University,
Beijing 100875, China}
\author{Gang Xiong}
\affiliation{Department of Physics, Beijing Normal University,
Beijing 100875, China}
\author{Ming Gong}
\affiliation{Department of Physics and Astronomy, Washington State University, 
Pullman, WA 99164 USA}
\author{Xin-Qi Li}
\email{lixinqi@bnu.edu.cn}
\affiliation{Department of Physics, Beijing Normal University,
Beijing 100875, China}

\date{\today}

\begin{abstract}
We consider an experimentally feasible setup to demonstrate
the existence and coherent dynamics of Majorana fermion.
The transport setup consists of a quantum dot and a tunnel-coupled
semiconductor nanowire which is anticipated
to generate Majorana excitations under some conditions.
For transport under finite bias voltage,
we find that a subtraction of the source and drain currents
can expose the essential feature of the Majorana fermion,
including the zero-energy nature by gate-voltage
modulating the dot level.
Moreover, coherent oscillating dynamics of the Majorana fermion
between the nanowire and the quantum dot is reflected
in the shot noise, via a spectral dip together with
a pronounced zero-frequency noise enhancement effect.
Important parameters, e.g., for the Majorana's mutual interaction
and its coupling to the quantum dot, can be extracted out
in experiment using the derived analytic results.
\end{abstract}


\maketitle

\graphicspath{{figure/}}


The Majorana fermions, proposed in 1937 by Majorana \cite{MF37},
are exotic particles since each Majorana fermion
is its own antiparticle \cite{Wil0910}.
Recently, the search for Majorana fermions
in solid states, as emerged quasiparticles (elementary excitations),
is attracting a great deal of attention
\cite{Kit01,Fu08,Ore10,Sau10,Lut10,Ali10,Sau12}.
In addition to being of interest for some fundamental physics
such as the non-Abelian statistics, the Majorana
fermions emerged in solid states are also a promising candidate for
fault-tolerant topological quantum computation (TQC)
\cite{Kit03,Ste10,Nay08,Ali11}.
In solid states, the predicted Majorana bound states (MBSs)
are coherent superpositions of electron and
hole excitations of zero energy, which can appear for instance in the
5/2 fractional quantum Hall system\cite{Moore1991}
and the $p$-wave superconductor and superfluid\cite{Read2000}.
In practice, an effective $p$-wave superconductor
can be realized by a semiconductor nanowire
with Rashba spin-orbit interaction and Zeeman splitting, and
in proximity to an $s$-wave superconductor \cite{Ore10,Sau10,Sau12}.

Of crucial importance is then a full experimental observation of
the Majorana fermion in solid states, from multi-angles in a variety
of contexts to demonstrate its existence and coherent dynamics.
Some existing proposals include analyzing the tunneling spectroscopy
which may reveal characteristic zero-bias conductance peak \cite{Liu11}
and peculiar noise behaviors \cite{Bol07,Law09},
verifying the nonlocality nature of the MBSs \cite{Nil08,LeeDH08},
and observing the 4$\pi$ periodic Majorana-Josephson currents
\cite{Kit01,Lut10,Ore10,Fu09}.
We notice that a breakthrough was achieved
in a most latest experiment \cite{Kou12}
where, using the tunneling spectroscopy,
a mid-gap Majorana state was detected,
based on the observation that a zero-bias-peak stickily appears
in the differential conductance through an InSb nanowire,
while the nanowire is in contact with a superconducting electrode (NbTiN).

In this work we consider a setup based on transport through
a quantum dot (QD) to probe the existence and coherent dynamics
of Majorana fermion, by weakly tunnel-coupling the QD to a nanowire
as realized in the above-mentioned experiment \cite{Kou12}.
We find that a subtraction of the source and drain currents
can expose the essential feature of the Majorana fermion,
and the zero-energy nature can be justified
by gate-voltage modulating the QD level.
Moreover, oscillating dynamics of the Majorana fermion
between the nanowire and the quantum dot can be identified
from the shot noise, via a spectral dip together with
a pronounced zero-frequency noise enhancement effect.

\begin{figure}[!htbp]
  \centering
  \includegraphics[width=6.5cm]{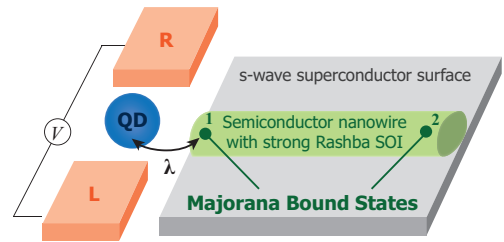}
\caption{(Color online)
Schematic setup for detecting Majorana fermion
by transport through a quantum dot,
while the dot is coupled to a semiconductor nanowire
which is in contact with an $s$-wave superconductor.
Under appropriate conditions such as with large
Zeeman splitting and strong spin-orbit interaction (SOI),
a pair of Majorana bound states is anticipated
to emerge at the ends of the nanowire. }\label{FIG1}
\end{figure}

%
{\it Model.}---
Figure 1 shows schematically the setup that will be studied in this work,
where the transport through the quantum dot (QD) is employed to
detect the Majorana fermion emerged in the tunnel-coupled nanowire.
Combining a strong Rashba spin-orbit interaction
and the Zeeman splitting, it was shown
that the proximity-effect-induced $s$-wave superconductivity in the nanowire
can support electron-hole quasiparticle excitations of Majorana bound states
(MBSs) at the ends of the nanowire \cite{Ore10,Sau10,Sau12}.
Since the Zeeman splitting should be large enough in order to
drive the wire into a topological superconducting phase,
we can assume it much larger than the transport bias voltage and
the tunnel coupling amplitude and rates.
In this case, we can model the QD by a single resonant level
and treat the electron as spinless particle.
Accordingly, the whole system of Fig.\ 1 can be modeled by
$H=H_{\rm Leads}+H_{\rm sys}+H_T$.
$H_{\rm Leads}=\sum_{\alp=L,R}\sum_{k} \vep_{\alp k}\cdg_{\alp k}c_{\alp k}$
describes the normal metallic leads,
$H_T=\sum_{\alp=L,R}\sum_{k}t_{\alp k} d \cdg_{\alp k}+{\rm H.c.}$
is for the tunneling between the leads and the dot,
and the low-energy effective Hamiltonian for the central system reads
\beq\label{ham1}
    H_{\rm sys}=\eps_D \ddg d+\frac{i}{2}\eps_M\gam_1\gam_2
    +( \lam d-\lam^* \ddg)\gam_1 .
\eeq
Here $\cdg_{\alp k}(c_{\alp k})$ and $\ddg(d)$ are, respectively,
the electron creation (annihilation) operators of the leads and the dot,
with the corresponding energies $\vep_{\alp k}$ and $\eps_D$.
Particularly, in \Eq{ham1}, the second term describes the paired MBSs
generated locally at the ends of the nanowire but with a mutual
coupling $\eps_M\sim e^{-\ell/\xi}$, where $\ell$ is the wire length
and $\xi$ is the superconducting coherent length.
The last term in \Eq{ham1} describes the tunnel coupling between the dot
and the left MBS.
For spinless dot level, we can choose a real constant for $\lam$, while
in general $\lam$ has a phase factor associated with the spin direction.


To solve the transport problem associated with \Eq{ham1},
it is convenient to switch from the Majorana representation
to the regular fermion one, through the exact transformation
$\gam_1=\fdg+f$ and $\gam_2=i(\fdg-f)$.
$f$ is the regular fermion operator, satisfying the
anti-commutative relation $\{f,\fdg\}=1$.
Accordingly, we rewrite $H_{\rm sys}$ as
\beq\label{ham2}
    H_{\rm sys}=\eps_D \ddg d+\eps_M\left(\fdg f-\frac{1}{2}\right)
    +\lam (d-\ddg)(\fdg+f)  .
\eeq
For latter convenience of discussion, we rearrange the
tunnel coupling term in \Eq{ham2} as
$H_1=(\lam \fdg d + \lam_1 \fdg\ddg ) + {\rm H.c.}$
where $\lam_1=\lam$ or $0$ corresponds to
coupling to a MBS or a regular fermion bound state.
In the transformed representation, the basis states of the central
system are given by $\ket{n_dn_f}$, with $n_d$ and $n_f$ being
0 or 1 so that we have four basis states
$\{\ket{00}, \ket{10}, \ket{01}, \ket{11}\}$.


Rather than the linear response\cite{Liu11},
we consider transport through the quantum dot under finite (large) bias
(in this regime the temperature can be approximated to zero
in calculating the transport currents).
This case corresponds to the single dot level deeply
buried into the voltage window, which allows us to employ
the standard Born-Markov master equation to solve the problem.
For the transport setup, the two leads are regarded as a generalized environment,
and the embedded system in between the leads as the {\it system of interest}
which is described by the reduced density matrix $\rho(t)$.
For the latter use of calculating the noise spectrum,
we present below the particle-number resolved version
of the master equation for $\rho^{(n)}(t)$ as\cite{Li05}
\beqn\label{rho1}
    \dot{\rho}^{(n)}&\!\!=\!\!&-i\LL \rho^{(n)}
    \!-\!\frac{\Gam_L}{2}\left( d \ddg\rho^{(n)}
      +\rho^{(n)} d \ddg-2\ddg\rho^{(n)}d \right)\re
    &&-\frac{\Gam_R}{2}\left(\ddg d \rho^{(n)}
      +\rho^{(n)}\ddg d-2d\rho^{(n-1)}\ddg\right) ,
\eeqn
with ``$n$" the electron number transferred through the central system
and satisfying $\sum_{n=0}^{\infty}\rho^{(n)}(t)=\rho(t)$.
Here we introduced the Liouvillian as
$\LL \rho\equiv [H_{\rm sys}, \rho]$,
and the tunneling rate of the dot to the lead as
$\Gam_\alp=2\pi g_\alp|t_\alp|^2$,
with $g_\alp$ the density of states of the lead $\alp$ ($L$ or $R$).
Corresponding to \Eq{rho1}, the {\it unconditional} Lindblad
master equation reads
$ \dot{\rho} = -i\LL \rho + \Gamma_L {\cal D}[d^{\dagger}]\rho
 + \Gamma_R {\cal D}[d]\rho $,
where ${\cal D}[A]\rho\equiv A\rho A^{\dagger}
-\frac{1}{2}\{A^{\dagger}A, \rho\}$.

%
{\it Current.}---First we consider the transport current,
which can be evaluated in a rather simple way as follows.
Instead of solving \Eq{rho1}, we evaluate the unconditional
Lindblad master equation for $\rho(t)$, and calculate the
average occupation of the quantum dot,
$\bar{n}_{d}=\tr\left[\ddg d \rho(t)\right]$.
Then, the left- and right-lead currents are obtained as
\beq
    I_L(t)=e\Gam_L (1-\bar{n}_{d}),
    ~~~ I_R(t)=e\Gam_R \bar{n}_{d} .
\eeq
For steady-state currents, analytic results read
\bsub
\beq
    I_L^s=\frac{e\Gam_L\Gam_R}{\Gam}
    \left[1+\frac{4(1/y-1)\lam^2}
    {\Gam^2+4(\eps_D^2+\eps_M^2+2\lam^2)}\right] ,
\eeq
\beq
    I_R^s=\frac{e\Gam_L\Gam_R}{\Gam}
    \left[1+\frac{4(y-1)\lam^2}
    {\Gam^2+4(\eps_D^2+\eps_M^2+2\lam^2)}\right] .
\eeq
\esub
Here we introduced the total rate $\Gam=\Gam_L+\Gam_R$,
and the asymmetric factor $y=\Gam_R/\Gam_L$.
Also, in the above results we restore $\lam_1=\lam$
as it should be for coupling to Majorana fermion.

We notice that the first term, $e\Gam_L\Gam_R/\Gam$,
is just the steady-state current through a single resonant level
without any side coupling, or coupling to a regular
fermion bound state.
In the latter case, the side coupling
does not affect the stationary state,
but only affects the transient behaviors (e.g., the noise properties).
The second term, in the $I_{L(R)}^s$, originates from the Majorana nature,
particularly its hole excitation (antiparticle component).
However, of interest is that this term appears as a consequence
of the interplay of the electron and hole excitations
in the Majorana fermion, but not their separate individual role.

\begin{figure}[!htbp]
  \centering
  \includegraphics[width=6cm]{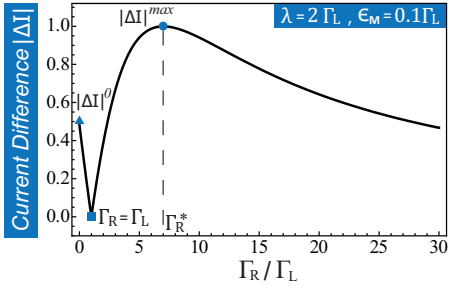}
\caption{(Color online)
Difference of the {\it steady-state} source and drain currents,
unlike the usual case that is zero,
which reveals the Majorana's unique nature of electron
and hole (i.e.,the antiparticle component) excitation.
This current difference depends on the ratio
of $\Gamma_L$ and $\Gamma_R$, which not only indicates
the existence (effect) of the Majorana fermion,
but also provides an access
to extract its coupling amplitude to the quantum dot.
In this plot we set $\eps_D=0$. }\label{FIG2}
\end{figure}

Moreover, an unusual feature appears that, even in steady state,
the source and drain currents are unequal to one another,
with a difference of
\beq\label{DI}
    \Del I=I_R^s-I_L^s=\frac{4e(\Gam_R-\Gam_L)\lam^2}
    {\Gam^2+4(\eps_D^2+\eps_M^2+2\lam^2)} .
\eeq
This ``non-conservation" of currents is originally owing to the presence
of superconductor condensate which can absorb and provide electrons.
In the emerged Majorana picture, it appears as a consequence
of the tunneling-induced creation and annihilation of electron pair
in the absence of bias voltage (across the quantum dot),
because of the superposition of electron and hole excitations.
Based on \Eq{DI}, one can verify the existence of the Majorana fermion
from two aspects.
{\it (i)} Modulating the quantum dot level by a gate voltage,
the current difference will exhibit a Lorentzian ``lineshape"
for the dot level ($\eps_D$) dependence,
which centers at the Majorana's zero energy.
{\it (ii)} As shown in Fig.\ 2, the current difference
depends on the asymmetry of $\Gamma_L$ and $\Gamma_R$.
In particular, we find that (only) for $\Gamma_R=\Gamma_L$
the source and drain currents are equal to each other.
And, there exists a special setup with the right rate $\Gam_R$
equal to $\Gam_R^*$,
\beq
    \Gam_R^*=\Gam_L+2\sqrt{\Gam_L^2+2\lam^2+\eps_D^2+\eps_M^2},
\eeq
the current difference reaches a maximum
\beq
    \Del I^{max}=\frac{2e\lam^2}{\Gam_R^*+\Gam_L}
\eeq
This simple result can also be used to justify the existence of
the Majorana fermion, and to extract its coupling amplitude ($\lam$)
to the quantum dot as well.

{\it Shot Noise.}---
As recognized in quantum transport, the shot noise can
carry valuable information beyond the steady-state current.
We therefore, in the following, further explore the shot noise
properties of our setup.
Specifically, we consider the current correlator
$S_\alp(t)=\frac{1}{2}\langle \{\delta I_\alp(t),
\delta I_\alp(0)\}\rangle$, with $\alp=L$ and $R$
and $\delta I_\alp(t)=I_\alp(t)-I^s_\alp$.
Here $I^s_\alp$ is the steady-state current in the $\alp$-th lead;
and the (quantum statistical) average in defining the correlator
is over the steady state as well.
The shot noise spectrum, i.e., the Fourier transform of $S_\alp(t)$,
can be calculated most conveniently within the number-resolved
master equation formalism using the McDonald's formula \cite{Li05}
\beq\label{Sw}
S_{\alp}(\ome)=2e^2\ome \int^\iny_0 d\tau \sin(\ome \tau) \frac{d}{d\tau}
    \left[\langle n_{\alp}^2(\tau)\rangle
     - \langle n_{\alp}(\tau)\rangle^2\right],
\eeq
where
$\langle n_{\alp}(\tau)\rangle\equiv \sum_{n_{\alp}}n_{\alp}
\rho^{(n_{\alp})}(\tau)$, and
$\langle n_{\alp}^2(\tau)\rangle\equiv \sum_{n_{\alp}}
n_{\alp}^2 \rho^{(n_{\alp})}(\tau)$. Here,
$\rho^{(n_{\alp})}(\tau)$ is defined by counting the electron number
``$n_{\alp}$" through the $\alp$-th junction
starting from an arbitrary moment in the steady state.
Therefore, we have $\langle n_{\alp}(\tau)\rangle=I^s_{\alp}\tau$.
To calculate $\langle n_{\alp}^2(\tau)\rangle$,
combining the number-resolved master equation,
we can construct its equation-of-motion (EOM).
Then, Laplace transforming the EOM and solving the resultant
algebraic equations can directly give the result of
$S_{\alp}(\ome)$, by noting that the integral in \Eq{Sw}
is actually a Laplace transformation since,
$\sin(\omega\tau)=\frac{1}{2i}(e^{i\omega\tau}-e^{-i\omega\tau})$.

\begin{figure}[!htbp]
  \centering
  \includegraphics[width=7cm]{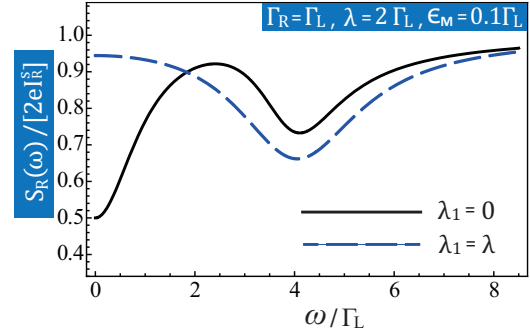}
  \caption{ (Color online)
Shot noise which reveals a spectral {\it dip}
and a zero-frequency {\it enhancement effect}.
The former reflects a coherent oscillation in the system
which indicates the formation of a bound state coupled
to the quantum dot, while the latter originates from
the peculiar nature of the Majorana excitation.
$\lambda_1=\lambda$ corresponds to coupling to
the Majorana bound state,
while $\lambda_1=0$ to a regular bound state.
As in Fig.\ 2, we set also $\eps_D=0$ in this figure.  }\label{FIG4}
\end{figure}

%

Figure 3 displays a representative result for the shot noise spectrum.
First, we notice that a spectral {\it dip} appears
at the frequency $\omega_c\simeq 2\lam$ (for $\eps_D=0$ and small $\eps_M$),
which reflects an existence
of coherent oscillations in the central system,
with the characteristic frequency $\omega_c$.
This is an important signature, since it indicates
the emergence of {\it bound states} at the ends of the wire
with discrete energy, gapped from other higher energy continuum.
At this point,
we should keep in mind that the quantum dot is coupled to a
nanowire, so in the usual case the wire states are extended and have
almost continuous energies, which cannot support coherent oscillations
as indicated by the spectral dip in Fig.\ 3.
As a comparison, we plot in Fig.\ 3 as well the result
for the same quantum dot and with the same strength ($\lambda$)
coupled to a regular bound state.
While an ``oscillation dip" appears similarly at the same frequency,
nevertheless the zero (and low) frequency noise
differs remarkably from the Majorana case.

To characterize the zero frequency noise, we employ
the well-known Fano Factor, $F_\alp=S_\alp(0)/2eI_\alp^s$.
For symmetric rates $\Gam_R=\Gam_L=\Gam_0$,
we find $F_L=F_R$ and denote the Fano factor simply by $F$.
Analytically, we obtain
\beq\label{DFF}
    F-F^{\rm (R)}\!=\frac{2\lam^2\lam_1^2}
    {(\Gam_0^2+\eps_M^2)(\lam^2+\lam_1^2)+4\lam^2\lam_1^2} .
\eeq
Here $F^{\rm (R)}$ is the Fano factor for
coupling to a regular (R) bound state.
Also, in this result we distinguish the coupling amplitudes
$\lam$ and $\lam_1$
(introduced previously in the coupling Hamiltonian).
We notice that $\lam$ and $\lam_1$
play identical (symmetric) role and
the difference, \Eq{DFF}, vanishes
if any of the amplitudes disappears.
We then understand that the zero-frequency noise enhancement
is arising from the peculiar nature of the Majorana excitation.
Therefore, the noise enhancement effect in Fig.\ 3
is another useful signature for Majorana excitation
at the ends of the nanowire.
Also, \Eq{DFF} can be used to extract the important parameters
of the Majorana's mutual interaction ($\eps_M$)
and its coupling with the quantum dot ($\lam$).
In particular, based on \Eq{DFF}, an even simpler result can be
obtained in the limit $\eps_M\rightarrow 0$ by noting
$\lam_1=\lam$ for Majorana fermion.
That is, $F=1-\frac{1}{2}[1+2(\lam/\Gam_0)^2]^{-1}$.
This result provides a very simple relation between the
Fano factor and the scaled coupling amplitude ($\lam/\Gam_0$).

Now we consider $\Gam_L\neq \Gam_R$.
Taking the limit $\eps_M\rightarrow 0$ and setting $\eps_D=0$,
we obtain
\bsub
\beq
    F_R=\frac{\Gam_L^2+\Gam_R^2}{\Gam^2}
    +\frac{8\Gam_R\Gam_L\lam^2\left[(5-3y)\Gam^2+16\lam^2\right]}
    {\Gam^2(\Gam^2+8\lam^2)^2}  ,
\eeq
\beq
    F_L=\frac{\Gam_L^2+\Gam_R^2}{\Gam^2}
    +\frac{8\Gam_R\Gam_L\lam^2\left[(5-3/y)\Gam^2+16\lam^2\right]}
    {\Gam^2(\Gam^2+8\lam^2)^2}  ,
\eeq
\esub
We notice that the first term in $F_L$ and $F_R$ is the Fano factor
corresponding to transport through an isolated
single-level quantum dot \cite{ChT91},
while the second term arises from coupling to Majorana fermion.
If $\Gam_L\neq \Gam_R$, we find
$F_R-F_L=24(\Gam_L-\Gam_R)\Gam\lam^2/(\Gam^2+8\lam^2)^2\neq 0$.
(This difference vanishes when $\Gam_L=\Gam_R$.)
As a final remark, we notice that the steady-state current through
the quantum dot cannot reveal the Majorana information
in the symmetric case ($\Gam_L=\Gam_R$).
However, the Fano factor given above carries such information,
since the second term in $F_{L(R)}$ does not vanish when $y=1$.
Note also that, if the quantum dot couples to a regular bound state
(another dot), the zero frequency noise (Fano factor) is the same
as the first term of the above $F_{L(R)}$,
being unaffected by the side coupling.

{\it Experimental Feasibility.}---
The semiconductor nanowire sketched in Fig.\ 1
can be the InSb wire utilized in the recent experiment \cite{Kou12}.
The InSb nanowire has a large $g$-factor ($g\simeq 50$),
and a strong Rashba-type
spin-orbit interaction (with energy $\sim$50 $\mu$eV).
For an applied magnetic field of $0.15$ Tesla,
the Zeeman splitting, $E_z\simeq 225~\mu{\rm eV}$,
starts to exceed the induced superconducting gap
$\Delta\simeq 200~\mu{\rm eV}$,
thus being able to support the
emergence of Majorana fermion at the ends of the InSb nanowire.
Therefore, a low temperature such as $T=100$ mK can
suppress the thermal excitation of the Majorana zero mode
to higher energy states.
As a consequence, the condition $k_BT<eV<E_z$ defines
a broad range of bias voltage across the quantum dot
to detect the Majorana's existence and its coherent dynamics.
For the latter purpose, one may tune the
tunneling rates $\Gamma_{L(R)}$ and the coupling energy $\lambda$
to the order of a few $\mu$eV, which can maintain
a coherent oscillation between the QD and the Majorana states,
provided the system has a coherence time longer than nanoseconds.
Moreover, the above parameters can guarantee a {\it large} bias
condition for electron transport through the quantum dot, which
is needed to ensure the employed master equation applicable.
The above estimates indicate that the proposed measurement scheme
in this work is feasible by the present state-of-the-art experiment.

%
{\it Summary.}---
We have proposed a quantum dot transport scheme to detect
the existence and coherent dynamics of Majorana fermion.
To be specific, we considered the quantum dot being
tunnel-coupled to a semiconductor nanowire which is anticipated
to generate Majorana excitations under appropriate conditions.
We carried out peculiar signatures in both the steady state current
and shot noise for our demonstrating purpose.

%
{\it Acknowledgments.}---
We thank Supeng Kou for stimulating discussions.
This work was supported by the NNSF of China (No. 101202101),
and the Major State Basic Research Project of China
(Nos.\ 2011CB808502 \& 2012CB932704).



\end{CJK*}
\end{document}